# Fast Planning in Stochastic Games

**Michael Kearns**
AT&T Labs
Florham Park, New Jersey
mkearns@research.att.com

**Yishay Mansour**
Tel Aviv University
Tel Aviv, Israel
mansour@math.tau.ac.il

**Satinder Singh**
AT&T Labs
Florham Park, New Jersey
baveja@research.att.com

## Abstract

Stochastic games generalize Markov decision processes (MDPs) to a multiagent setting by allowing the state transitions to depend jointly on all player actions, and having rewards determined by multiplayer matrix games at each state. We consider the problem of computing Nash equilibria in stochastic games, the analogue of planning in MDPs. We begin by providing a simple generalization of finite-horizon value iteration that computes a Nash strategy for each player in general-sum stochastic games. The algorithm takes an arbitrary *Nash selection function* as input, which allows the translation of local choices between multiple Nash equilibria into the selection of a single global Nash equilibrium.

Our main technical result is an algorithm for computing near-Nash equilibria in large or infinite state spaces. This algorithm builds on our finite-horizon value iteration algorithm, and adapts the sparse sampling methods of Kearns, Mansour and Ng (1999) to stochastic games. We conclude by describing a counterexample showing that infinite-horizon discounted value iteration, which was shown by Shapley to converge in the zero-sum case (a result we give extend slightly here), does not converge in the general-sum case.

## 1 INTRODUCTION

There has been increasing interest in artificial intelligence in multi-agent systems and problems. Fueled by the expanding potential for large-scale populations of autonomous programs in areas as diverse as online trading and auctions, personalized web crawling, and many other areas, such work seeks to provide both algorithms and mathematical foundations for agents in complex, distributed environments.

Given the detailed theoretical and practical understanding of single-agent planning and learning in Markov decision processes (MDPs) that has been built over the last decade, one natural line of research is the extension of these algorithms and analyses to a multi-agent setting (Boutilier, Goldszmidt, and Sabata 1999; Brafman and Tennenholtz 1999; Hu and Wellman 1998). The work presented here is a contribution to this line. We consider the problem of computing Nash equilibria in stochastic games. Stochastic games generalize MDPs to a multi-agent setting by allowing the state transitions to depend jointly on all agent actions, and having the immediate rewards at each state determined by a multi-agent general-sum matrix game associated with the state. If we view the computation of an optimal policy in a given MDP as the problem of *planning* for a single agent's interaction with its fixed stochastic environment, the natural analogue in a stochastic game would be the computation of a Nash equilibrium, which allows all agents to simultaneously enjoy a best response policy to the others.

We begin by providing a generalization of finite-horizon value iteration in MDPs that computes, in time polynomial in the number of states, a Nash strategy for each player in any given two-player [1], general-sum stochastic game. We introduce the notion of a *Nash selection function*, which is simply a way of extracting *local* Nash equilibria from the matrix games stored at each state. The key observation behind our value iteration generalization is that local application of *any* Nash selection function to the appropriate backup matrices will yield a global Nash equilibrium for the stochastic game. Different global so-lutions can be found by varying the underlying Nash selection function used by the algorithm.

---

[1] For simplicity, we present all of our results for the two-player case. They can be generalized to the $k$-player case at the cost of a factor exponential in $k$, which cannot be avoided without special assumptions on the game.

We then proceed to present our central technical contribution, which is an online *sparse sampling* algorithm for computing near-Nash equilibria in general-sum stochastic games. In the same way that recent work of Kearns, Mansour and Ng (1999) provided an online planning algorithm for large or infinite MDPs, we build on our finite-horizon value iteration and give a randomized algorithm for computing each half of a Nash policy in a stochastic game for which we have only a *generative model* (the ability to sample the state transition distributions and rewards). Like the sparse sampling algorithm of Kearns et al. for the MDP case, the per-state running time of our algorithm has *no dependence* on the state space size, but is exponential in the horizon time. This provides a rather different trade-off than our basic value iteration algorithm, whose dependence on the state space size is quadratic, but whose dependence on the horizon time is linear. Like the earlier work, one can view the new algorithm as greatly reducing the *width* or degree (from linear in state space size to constant) of a full look-ahead tree, while leaving the depth unchanged.

The strong guarantee we prove for our algorithm is that it computes a near-Nash equilibrium in any stochastic game: as long as one player chooses to play their half of the output of the algorithm at each state, the other player provably has little incentive to deviate from playing their half. Since our algorithm is randomized (due to the sparse sampling), and because of the potential sensitivity of Nash equilibria to small fluctuations in the game, it is important that both players play according to a common copy of the algorithm. (Technically, this is known as a *correlated* equilibrium.) Our algorithm can thus be viewed as providing a centralized solution for the players that is near-Nash.

We close with a discussion of value iteration in the infinite horizon discounted case. We give a small generalization of Shapley's proof of convergence in the zero-sum case, but then provide a rather strong counterexample to convergence in the general sum case, highlighting an interesting difference between planning in MDPs and planning in stochastic games.

## 2 DEFINITIONS AND NOTATION

We adopt standard terminology and notation from classical game theory. A *two-player game* is defined by a matrix pair $(M_1, M_2)$, specifying the *payoffs* for the row player (player 1) and the column player (player 2), respectively. We shall assume without loss of generality throughout the paper that all game matrices $M_1$ and $M_2$ are $n$ by $n$, and that their indices thus range from 1 to $n$. If the row player chooses the index (or *pure strategy*) $i$ and the column player chooses the index $j$, the former receives payoff $M_1(i,j)$ and the latter $M_2(i,j)$. More generally, if $\alpha$ and $\beta$ are distributions (or *mixed strategies*) over the row and column indices, the expected payoff to player $k \in \{1,2\}$ is $M_k(\alpha, \beta) \doteq \mathbf{E}_{i \in \alpha, j \in \beta}[M_k(i,j)]$, where the notation $i \in \alpha$ indicates that $i$ is distributed according to $\alpha$. We say that the game $(M_1, M_2)$ is *zero-sum* if $M_2 = -M_1$.

The mixed strategy pair $(\alpha, \beta)$ is said to be a *Nash equilibrium* (or *Nash pair*) for the game $(M_1, M_2)$ if (i) for any mixed strategy $\alpha'$, $M_1(\alpha', \beta) \leq M_1(\alpha, \beta)$, and (ii) for any mixed strategy $\beta'$, $M_2(\alpha, \beta') \leq M_2(\alpha, \beta)$. In other words, as long as one player plays their half of the Nash pair, the other player has no incentive to switch from their half of the Nash pair. We will also need the standard notion of approximate Nash equilibria. Thus, we say that $(\alpha, \beta)$ is $\epsilon$-*Nash* for $(M_1, M_2)$ if (i) for any mixed strategy $\alpha'$, $M_1(\alpha', \beta) \leq M_1(\alpha, \beta) + \epsilon$, and (ii) for any mixed strategy $\beta'$, $M_2(\alpha, \beta') \leq M_2(\alpha, \beta) + \epsilon$.

It is well-known (Owen 1995) that every game has at least one Nash pair in the space of mixed (but not necessarily pure) strategies, and many games have multiple Nash pairs. Furthermore, in the zero-sum case, if $(\alpha, \beta)$ and $(\alpha', \beta')$ are both Nash pairs, then $M_1(\alpha, \beta) = M_1(\alpha', \beta')$, and $(\alpha, \beta')$ and $(\alpha', \beta)$ are also Nash pairs. In other words, in the zero-sum case, the payoff to the players is always the same under any Nash pair, and Nash pairs can be freely "mixed" together to form new Nash pairs. Since in the zero-sum case the payoff to the column player is always the negative of the row player, we can unambiguously refer to the *value* of the game $v(M_1) \doteq M_1(\alpha, \beta)$, where $(\alpha, \beta)$ is any Nash pair. In the general-sum case, different Nash pairs may yield different payoffs to the players, and in general Nash pairs cannot be mixed in the manner described.

The *security level* for the row player is defined as $s_1(M_1, M_2) = s_1(M_1) \doteq \max_\alpha \min_\beta M_1(\alpha, \beta)$ and for the column player as $s_2(M_1, M_2) = s_2(M_2) \doteq \max_\beta \min_\alpha M_2(\alpha, \beta)$. A mixed strategy achieving the maximum in each case is referred to as a *security strategy* for the player. The security level for a player is the payoff that player can ensure for themselves regardless of their opponent's behavior. In the zero-sum case, the notion of Nash pair coincides with that of security pair, and the value of a game is simply the security level for the row player; in the general case, however, the security level of a player may be lower than their payoff in a Nash pair.

A *stochastic game* $G$ over a state space $S$ consists of a designated *start state* $s_0 \in S$, a matrix game $(M_1[s], M_2[s])$ for every state $s \in S$, and *transition probabilities* $P(s'|s,i,j)$ for every $s, s' \in S$, every pure row strategy $i$, and every pure column strategy $j$. Each step of play in a stochastic game proceeds as follows. If play is currently in state $s$ and the two players play mixed strategies $\alpha$ and $\beta$, then pure strategies $i$ and $j$ are chosen according to $\alpha$ and $\beta$ respectively, the players receive *immediate payoffs* $M_1[s](i,j)$ and $M_2[s](i,j)$ (and thus have expected immediate payoffs $M_1[s](\alpha,\beta)$ and $M_2[s](\alpha,\beta)$), and the next state $s'$ is drawn according to the transition probabilities $P(\cdot|s,i,j)$. Thus both the immediate payoffs to the players *and* the state transition depend on the actions of *both* players. If the game matrices at every state of $G$ are zero-sum we say that $G$ is zero-sum.

We shall consider two different standard measures of the overall *total return* received by the players. In the *infinite-horizon discounted* case, play begins at $s_0$ and proceeds forever; if a player receives payoffs $r_0, r_1, r_2, \ldots$, they are credited with total return $r_0 + \gamma r_1 + \gamma^2 r_2 + \cdots$, where $0 \le \gamma < 1$ is the *discount factor*. In the *finite-horizon undiscounted* (also called the *T-step average*) case, which is our main interest, play begins at $s_0$ and proceeds for exactly $T$ steps; if a player receives payoffs $r_0, r_1, r_2, \ldots, r_{T-1}$ they are credited with total return $(1/T)(r_0 + r_1 + \cdots r_{T-1})$. We shall use $R_{\max}$ to denote the largest absolute value of any element in the matrices $M_k[s]$. For ease of exposition we will often make the assumption that $R_{\max} \le 1$ without loss of generality. The goal of each player in a stochastic game is to maximize their expected total return from the designated start state.

A *policy* for a player in a stochastic game $G$ is a mapping $\pi(s)$ from states $s \in S$ to mixed strategies to be played at the matrix game at $s$. A *time-dependent policy* $\pi(s,t)$ allows the mixed strategy chosen to depend on the number $t$ of steps *remaining* in the $T$-step game. It is known (Owen 1995) that we can restrict attention to policies (in the infinite-horizon discounted case) or time-dependent policies (in the finite-horizon case) without loss of generality — that is, no advantage can be gained by a player by considering the history of play. If $\pi_1$ and $\pi_2$ are policies in a matrix game $G$ with designated start state $s_0$, we use $G_k(s_0, \pi_1, \pi_2)$, $k \in \{1,2\}$, to denote the expected infinite-horizon discounted return to player $k$, and $G_k(T, s_0, \pi_1, \pi_2)$ to denote the expected $T$-step average return. The notion of Nash pairs extends naturally to stochastic games: we say that $(\pi_1, \pi_2)$ is a Nash pair if for any start state $s_0$ and any $\pi_1'$, $G_1(s_0, \pi_1', \pi_2) \le G_1(s_0, \pi_1, \pi_2)$, and for any start state $s_0$ and any $\pi_2'$, $G_2(s_0, \pi_1, \pi_2') \le G_2(s_0, \pi_1, \pi_2)$ (with the obvious generalization for the $T$-step case, and the obvious generalization to the notion of $\epsilon$-Nash). Again, it is known (Owen 1995) that for the infinite discounted case, a Nash pair always exists in the space of policies, and for the average return case, a Nash pair always exists in the space of time-dependent policies. The notions of security level and security policies in a stochastic game are defined analogously. The subject of this paper is the computation of Nash and $\epsilon$-Nash policy pairs in stochastic games, including stochastic games with large or infinite state spaces.

Since the matrix game at any given state in a stochastic game may have many Nash equilibria, it is easy to see that there may be exponentially many Nash equilibria in policy space for a stochastic game. One question we face in this paper is how to turn *local* decisions at each state into a global Nash or near-Nash policy pair. For this we introduce the notion of a *Nash selection function*. For any matrix game $(M_1, M_2)$, a Nash selection function $f$ returns a pair of mixed strategies $f(M_1, M_2) = (\alpha, \beta)$ that are a Nash pair for $(M_1, M_2)$. We denote the payoff to the row player under this Nash pair by $v_f^1(M_1, M_2) \doteq M_1(f(M_1, M_2))$ and the payoff to the column player by $v_f^2(M_1, M_2) \doteq M_2(f(M_1, M_2))$. Thus, a Nash selection function — which we allow to be arbitrary in most of our results — is simply a specific way of making choices of how to behave in isolated matrix games. Looking ahead slightly, we will show how the application of any Nash selection function to the appropriate backup matrices yields a Nash policy pair for the global stochastic game.

Similarly, a *security selection function* $f$ returns a pair of mixed strategies $f(M_1, M_2) = (\alpha, \beta)$ that are a security pair for $(M_1, M_2)$. In this case we use $v_f^k(M_1, M_2)$ to denote the security level of player $k$ in the game $(M_1, M_2)$. For both Nash and security selection functions, we denote the row and column strategies returned by $f$ by $f_1(M_1, M_2)$ and $f_2(M_1, M_2)$, respectively.

## 3 FINITE VALUE ITERATION IN GENERAL STOCHASTIC GAMES

We begin by describing a *finite-horizon, undiscounted value iteration* algorithm for general-sum stochastic games, given in Figure 1. This algorithm and its analysis are a straightforward generalization of classical finite-horizon value iteration in Markov decision processes, but form a necessary basis for our large state-space algorithm and analysis.

The algorithm outputs a pair of time-dependent policies $\pi_k(s,t)$, $k \in \{1,2\}$, mapping any state $s$ and the time remaining $t \le T$ to a mixed strategy for player $k$. The algorithm assumes access to an arbi-

---

**Algorithm FiniteVI($T$):**
  **Initialization:**
    For all $s \in S$, $k \in 1, 2$:
      $Q_k[s, 0] \leftarrow M_k[s]$;
      $\pi_k(s, 0) \leftarrow f_k(M_1[s], M_2[s])$;
  **Iteration $t = 1 \ldots T$:**
    For all $s \in S$, $k \in \{1, 2\}$:
      For all pure strategies $i$ and $j$:
        $Q_k[s,t](i,j) \leftarrow M_k[s](i,j) + \sum_{s'} P(s'|s,i,j) v_f^k(Q_1[s', t-1], Q_2[s', t-1])$;
      $\pi_k(s, t) \leftarrow f_k(Q_1[s,t], Q_2[s,t])$;
  Return the policy pair $(\pi_1, \pi_2)$;

---

Figure 1: Algorithm **FiniteVI** for computing Nash equilibria in finite-horizon undiscounted stochastic games. Recall that $f$ is an arbitrary Nash selection function, $f_k$ extracts the mixed strategy selected by $f$ for player $k$, and $v_f^k$ extracts the value of the selected Nash to player $k$.

trary, fixed stochastic game $G$ (specifying the payoff matrices $(M_1[s], M_s[s])$ and the transition probabilities $P(s'|s, i, j)$ for every state), and an arbitrary, fixed Nash selection function $f$. We will show that the algorithm outputs a pair of policies that are a Nash equilibrium for the $T$-step stochastic game from any start state. The running time is quadratic in the number of states.

The algorithm must maintain backup matrices at each step rather than just backup values. The main observation behind the algorithm is the fact that choices between the many global Nash equilibria for the $T$-step stochastic game can be implicitly made by application of an *arbitrary* Nash selection function. The algorithm also bears similarity to the infinite-horizon discounted value iteration algorithm studied by Shapley (1953), which we revisit in Section 5. While Shapley proved that the infinite-horizon discounted algorithm converges to a Nash equilibrium in the *zero-sum* case (and indeed, as we shall show, does not converge in the general-sum case), the main result of this section shows that our finite-horizon algorithm converges to Nash in the *general-sum* setting.

**Theorem 1** *Let $G$ be any stochastic game, and let $f$ be any Nash selection function. Then the policy pair $\pi_k(s, t)$ output by* **FiniteVI**($f$) *is a Nash pair for the $T$-step stochastic game $G$ from any start state.*

**Proof:** The proof is by induction on the number of steps $T$. The base case $T = 0$ is straightforward. If the start state is $s_0$, and this is the only step of play, then clearly the players should play a Nash pair for the game $(M_1[s_0], M_2[s_0])$ defined at $s_0$. Since the initialization step given in algorithm **FiniteVI** explicitly specifies that the $\pi_k(s_0, 0)$ are simply the Nash pair identified by the Nash selection function $f$, this will be satisfied.

Now suppose the theorem holds for all $T < T_0$. Let us fix the policy played by player 2 to be $\pi_2(s, t)$ for all states $s$ and time remaining $t \leq T_0$, and consider whether player 1 could benefit by deviating from $\pi_2(s, t)$. By the inductive hypothesis, player 1 cannot benefit by deviating at any point after the first step of play at $s_0$. At the first step of play, since player 2 is playing according to $\pi_2(s_0, T_0)$, he will play his half of the Nash pair for the game $(Q_1[s_0, T_0], Q_2[s_0, T_0])$ chosen by the Nash selection function $f$. Furthermore, since play after the first step must be the Nash pair computed by the algorithm for the $T_0 - 1$ step game, the matrices $Q_k[s_0, T_0]$ contain the true total average return received by the players under any choice of initial actions at $s_0$. Therefore player 1 cannot benefit by deviating from the choice of action dictated by $\pi(s_0, T_0)$, and the theorem is proved. □(Theorem 1)

## 4 A SPARSE SAMPLING ALGORITHM IN LARGE GAMES

Algorithm **FiniteVI** computes full expectations over next states in order to compute the backup matrices $\hat{Q}_k[s, t]$, and thus has a running time that scales quadratically with the number of states in the stochastic game. In contrast, in this section we present a *sparse sampling* algorithm for *on-line* planning in stochastic games. The per-state running time of this algorithm has *no dependence* on the state space size, and thus can be applied even to infinite state spaces, but it depends exponentially on the horizon time $T$. This is exactly the trade-off examined in the work of Kearns, Mansour and Ng (1999) (see also McAllester and Singh (1999)), who gave an on-line algorithm for the simpler problem of computing near-optimal policies in large Markov decision processes. The algorithm presented here can be viewed as a generalization of algorithm **FiniteVI** that extends the sparse sampling methods of these earlier works to the problem of com-puting Nash equilibria in large or infinite state spaces.

Algorithm **SparseGame**, which is presented in Figure 2, takes as input any state $s$ and time $T$. It assumes access to an arbitrary, fixed Nash selection function $f$. Rather than directly accessing full transition distributions for the underlying stochastic game $G$, the algorithm only assumes access to the immediate payoff matrices $M_k[s]$, and the ability to sample $P(\cdot|s, i, j)$ for any $(i, j)$. This is sometimes referred to as a *generative model* or simulator for the stochastic game. The algorithm returns a mixed strategy pair $(\alpha, \beta)$ to be played at $s$, along with values $\hat{Q}_1$ and $\hat{Q}_2$, whose properties we will discuss shortly.

The only aspect of the algorithm that has been left unspecified is the sample size $m$ of next states drawn. Note that due to the recursion, a call to **SparseGame** will build a recursive tree of size $m^T$, so the overall running time of the algorithm will be exponential in $T$. The question of interest is how large we must make $m$.

Algorithm **SparseGame** is an *on-line* or *local* algorithm in the sense that it takes a single state $s$ and the amount of time remaining $T$, and produces a mixed strategy pair $(\alpha, \beta)$. This algorithm, however, defines an obvious (global) *policy* for player $k$ in the stochastic game: namely, upon arriving in any state $s$ with $T$ steps left in the finite-horizon game, play the mixed strategy for player $k$ output by **SparseGame**$(s, T)$. Our goal is to prove that for a choice of $m$ with *no dependence* on the state space size, these global policies are near-Nash — that is, if one player chooses to always play their half of the mixed strategy pairs output by the algorithm at each state visited, then the other player cannot benefit much by deviating from the same policy. This result will be derived via a series of technical lemmas. We begin with some necessary definitions.

For any $s$ and $T$, we define $V_k[s, T]$ to be the expected return to player $k$ in the $T$-step game starting at $s$ when *both* players play according to **SparseGame** at every step. For any pure strategy pair $(i, j)$, we define $V_k[s, T](i, j)$ to be the expected return to player $k$ in the $T$-step game starting at $s$ when the first step play at $s$ is restricted to be $(i, j)$, and both players play according to **SparseGame** at every step afterwards. It will be fruitful to view the values $\hat{Q}_k$ returned by the call **SparseGame**$(s, T)$ as approximations to the $V_k[s, T]$, and to view the related matrices $\hat{Q}_k[s, T](i, j)$ computed by this call as approximations to the matrices $V_k[s, T](i, j)$. This view will be formally supported by a lemma shortly.

We define $B_k[s, T]$ to be the expected return to player $k$ when we fix their opponent to play according to **SparseGame** at every step, and player $k$ plays the *best response* policy — that is, the policy that maximizes their return given that their opponent plays according to **SparseGame**. (It is not important that we be able to compute this best response, since it is merely an artifact of the analysis.) Similarly, we define $B_k[s, T](i, j)$ to be the expected return to player $k$ when the first step of play at $s$ is restricted to be $(i, j)$, we fix the opponent of player $k$ to play according to **SparseGame** afterwards, player $k$ plays the best response policy afterwards.

In this terminology, the main theorem we would like to prove asserts that for any $s$ and $T$, the $V_k[s, T]$ (the values the players will receive by following the algorithm) are near the $B_k[s, T]$ (the values the players could obtain by deviating from the algorithm if their opponent follows the algorithm), for a sufficiently large sample size that is independent of the state space size.

**Theorem 2** *For any $s$ and $T$, and for any $\epsilon > 0$, provided the sample size $m$ in algorithm **SparseGame** obeys*

$$m > c\left((T^3/\epsilon^2)\log(T/\epsilon) + T\log(n/\epsilon)\right)$$

*for some constant $c$ (where $n$ is the number of pure strategies available to both players at any state), then $|B_k[s, T] - V_k[s, T]| \leq 2T\epsilon$. In other words, the policy of always playing according to **SparseGame** is $2T\epsilon$-Nash for the $T$-step game starting at $s$.*

**Proof:** All expectations in the proof are over all the random samples generated by the call **SparseGame**$(s, T)$ (including the random samples generated by the recursive calls). We begin by noting:

$$\begin{aligned}|B_k[s,T] - V_k[s,T]| &= \mathbf{E}\left[|B_k[s,T] - V_k[s,T]|\right] \\ &\leq \mathbf{E}\left[|B_k[s,T] - \hat{Q}_k[s,T]|\right] + \\ &\quad \mathbf{E}\left[|\hat{Q}_k[s,T] - V_k[s,T]|\right]\end{aligned}$$

by the triangle inequality and linearity of expectation. We now state a lemma bounding the second expectation in this sum, and also bounding a related quantity that we will need later.

**Lemma 3** *For any $s$ and $T$, and for any $\epsilon > 0$, provided the sample size $m$ in algorithm **SparseGame** obeys*

$$m > c\left((T^3/\epsilon^2)\log(T/\epsilon) + T\log(n/\epsilon)\right)$$

*for some constant $c$ (where $n$ is the number of pure strategies available to both players at any state), then*

$$\mathbf{E}\left[|\hat{Q}_k[s,T] - V_k[s,T]|\right] \leq \epsilon$$

```
Algorithm SparseGame(s, T):
    If T = 0 (base case):
        α ← f_1(M_1[s], M_2[s]);
        β ← f_2(M_1[s], M_2[s]);
        For k ∈ {1, 2}, Q̂_k ← M_k[s](α, β);
        Return (α, β, Q̂_1, Q̂_2);
    /* Else need to recurse below */
    For each pure strategy pair (i, j):
        Sample s'_1, ..., s'_m from P(·|s, i, j);
        For ℓ = 1, ..., m:
            (α', β', Q̂_1[s'_ℓ, T-1], Q̂_2[s'_ℓ, T-1]) ← SparseGame(s'_ℓ, T-1);
        For k ∈ {1, 2}, Q̂_k[s, T](i, j) ← M_k[s] + (1/m) Σ_{ℓ=1}^m Q̂_k[s'_ℓ, T-1];
    α ← f_1(Q̂_1[s, T], Q̂_2[s, T]);
    β ← f_2(Q̂_1[s, T], Q̂_2[s, T]);
    For k ∈ {1, 2}, Q̂_k = Q̂_k[s, T](α, β);
    Return (α, β, Q̂_1, Q̂_2);
```

Figure 2: Algorithm **SparseGame** for computing one step of an approximate Nash equilibrium in a large stochastic game. The algorithm assumes only access to a generative model of the game. Recall that $f$ is an arbitrary Nash selection function, $f_k$ extracts the mixed strategy selected by $f$ for player $k$, and $v_f^k$ extracts the value of the selected Nash to player $k$.

and

$$\mathbf{E}\left[\max_{(i,j)}\{|V_k[s,T](i,j) - \hat{Q}_k[s,T](i,j)|\}\right] \leq \epsilon.$$

**Proof:** By induction on $T$, we show that the probability that $|\hat{Q}_k[s,T] - V_k[s,T]| \geq \lambda T$ is bounded by some quantity $\delta_T$, and that for any fixed $(i,j)$, the probability that $|\hat{Q}_k[s,T](i,j) - V_k[s,T](i,j)| \geq \lambda T$ is bounded by some quantity $\delta'_T$. These bounds will then be used to bound the expectations of interest.

For the base case $T = 0$, we have $\delta_T = \delta'_T = 0$. Assume the inductive hypothesis holds for horizon times $1, \ldots, T-1$. We define the random variable $U_k[s,T](i,j) \doteq (1/m)\sum_{\ell=1}^m V_k[s'_\ell, T-1]$, where the $s'_\ell$ are the states sampled in the call **SparseGame**$(s,T)$. By the triangle inequality we may write

$$|\hat{Q}_k[s,T](i,j) - V_k[s,T](i,j)|$$
$$\leq |\hat{Q}_k[s,T](i,j) - U_k[s,T](i,j)| +$$
$$|U_k[s,T](i,j) - V_k[s,T](i,j)|.$$

For the second term in this sum, the probability of the event $|U_k[s,T](i,j) - V_k[s,T](i,j)| \geq \lambda$ can be bounded by $e^{-\lambda^2 m}$ via a standard Chernoff bound analysis. In order to bound the first term in the sum, we note that by definition,

$$\hat{Q}_k[s,T](i,j) - U_k[s,T](i,j) =$$
$$(1/m)\sum_{\ell=1}^m \hat{Q}_k[s'_\ell, T-1] - V_k[s'_\ell, T-1].$$

By the inductive hypothesis, for each state $s'_\ell$, we have

$$|\hat{Q}_k[s'_\ell, T-1] - V_k[s'_\ell, T-1]| \leq \lambda(T-1)$$

with probability $1 - \delta_{T-1}$. This implies that with probability $1 - m\delta_{T-1}$, the bound applies to all the sampled states $s'_\ell$. In such a case we have that for any fixed $(i,j)$

$$|\hat{Q}_k[s,T](i,j) - U_k[s,T](i,j)| \leq \lambda(T-1).$$

Therefore, with probability

$$1 - \delta'_T \doteq 1 - (m\delta_{T-1} + e^{-\lambda^2 m})$$

we have that

$$|\hat{Q}_k[s,T](i,j) - V_k[s,T](i,j)| \leq \lambda T.$$

This completes the proof for the second half of the inductive hypothesis. For the first half of the inductive hypothesis we have

$$|\hat{Q}_k[s,T] - V_k[s,T]| \leq \max_{(i,j)} |\hat{Q}_k[s,T](i,j) - V_k[s,T](i,j)|.$$

This implies that with probability $1 - \delta_T \doteq 1 - n^2 \delta'_T$ we have

$$|\hat{Q}_k[s,T] - V_k[s,T]| \leq \lambda T$$

which completes the proof of the first half of the inductive hypothesis.

We would like to use the inductive hypothesis to prove the lemma. By the definition of $\delta_T$ we have

$$\mathbf{E}\left[|\hat{Q}_k[s,T] - V_k[s,T]|\right] \leq \delta_T + \lambda T.$$

Here we have used the assumption that $R_{\max} \leq 1$. Now we need to solve for $\delta_T$ in the recurrence set up above:

$$\delta_T = n^2 \delta'_T = n^2(m\delta_{T-1} + e^{-\lambda^2 m}) \leq (n^2 m)^T n^2 e^{-\lambda^2 m}.$$

To complete the proof we set $\lambda = \epsilon/2T$ and

$$m > (2T/\lambda^2)\log(T/\lambda^2) + T\log(2n^2/\epsilon) + 2\log(n).$$

Thus

$$m = O((T^3/\epsilon^2)\log(T/\epsilon) + T\log(n/\epsilon)).$$

The second bound of the lemma follows since with probability at most $n^2 \delta'_T$ we have

$$\max_{(i,j)}\{|V_k[s,T](i,j) - \hat{Q}_k[s,T](i,j)|\} \geq \lambda T$$

which completes the proof of the lemma. □(Lemma 3)

Returning to the main development, we have incurred an approximation cost $\epsilon$ by applying Lemma 3, and in exchange have reduced the task of bounding $|B_k[s,T] - V_k[s,T]|$ to the problem of bounding $\mathbf{E}\left[|B_k[s,T] - \hat{Q}_k[s,T]|\right]$. For this we need the following general lemma.

**Lemma 4** Let $(M_1, M_2)$ and $(\hat{M}_1, \hat{M}_2)$ be general-sum matrix games such that for all pure strategy pairs $(i,j)$, $|M_k(i,j) - \hat{M}_k(i,j)| \leq \Delta$ for $k \in \{1, 2\}$. Then if the mixed strategy pair $(\alpha, \beta)$ is a Nash in $(\hat{M}_1, \hat{M}_2)$, for any mixed strategy pair $(\alpha', \beta)$ we have $M_1(\alpha', \beta) - \hat{M}_1(\alpha, \beta) \leq \Delta$.

**Proof:** It is easy to see that for any $(\alpha, \beta)$ and $k \in \{1,2\}$, $|M_k(\alpha,\beta) - \hat{M}_k(\alpha,\beta)| \leq \Delta$. Now fix $(\alpha, \beta)$ to be some Nash pair in $(\hat{M}_1, \hat{M}_2)$. This implies that for any $\alpha'$ we have $\hat{M}_1(\alpha', \beta) \leq \hat{M}_1(\alpha, \beta)$. Combining the two observations we have that

$$M_1(\alpha', \beta) \leq \hat{M}_1(\alpha', \beta) + \Delta$$
$$\leq \hat{M}_1(\alpha, \beta) + \Delta.$$

□(Lemma 4)

Recall that by definition of **SparseGame**, $\hat{Q}_k[s,T]$ is obtained by computing a Nash equilibrium $(\alpha, \beta)$ (determined by the fixed Nash selection function) of the matrices $\hat{Q}_k[s,T](i,j)$, and setting (for example) $\hat{Q}_1[s,T] = \hat{Q}_1[s,T](\alpha,\beta)$. Application of Lemma 4 now lets us write

$$\mathbf{E}\left[|B_k[s,T] - \hat{Q}_k[s,T]|\right]$$
$$\leq \mathbf{E}\left[\max_{(i,j)}\{|B_k[s,T](i,j) - \hat{Q}_k[s,T](i,j)|\}\right]$$
$$\leq \mathbf{E}\left[\max_{(i,j)}\{|B_k[s,T](i,j) - V_k[s,T](i,j)|\}\right]$$
$$+ \mathbf{E}\left[\max_{(i,j)}\{|V_k[s,T](i,j) - \hat{Q}_k[s,T](i,j)|\}\right].$$

The second expectation in this sum, by Lemma 3, is bounded by $\epsilon$, incurring our second $\epsilon$ approximation cost. The first expectation no longer involves any random variables, so we may remove the expectation and concentrate on bounding

$$\max_{(i,j)}\{|B_k[s,T](i,j) - V_k[s,T](i,j)|\}.$$

For each fixed $(i,j)$, the difference

$$|B_k[s,T](i,j) - V_k[s,T](i,j)|$$

is just like the quantity we are attempting to bound in the theorem, but now we have fixed the first step of play to be $(i,j)$ and there are only $T-1$ steps afterwards. Thus we may apply a simple inductive argument:

$$|B_k[s,T](i,j) - V_k[s,T](i,j)|$$
$$= \left|\sum_{s'} P(s'|s,i,j)(B_k[s',T-1] - V_k[s',T-1])\right|$$
$$\leq \max_{s'}\{|B_k[s',T-1] - V_k[s',T-1]|\}.$$

This implies that we have

$$\max_s\{|B_k[s,T] - V_k[s,T]|\}$$
$$\leq 2\epsilon + \max_s\{|B_k[s,T-1] - V_k[s,T-1]|\}$$
$$\leq 2T\epsilon.$$

□(Theorem 2)

To apply Theorem 2, note that if we set $\epsilon = \epsilon'/T$ in the statement of the theorem and compute the resulting $m$ (which is polynomial in $T$ and $1/\epsilon'$), then the policies computed by the algorithm will be $\epsilon'$-Nash for the $T$-step stochastic game, and as already emphasized, the total per-state running time will be $m^T$. As mentioned in the introduction, it is important that the players play their respective halves of a *common* copy of the algorithm (a correlated equilibrium); playing an independent copy may not be a Nash strategy in some games. Briefly, the reason for this is that small variations in sampling may result in instabilities in the backup matrices computed. This instability is not a problem as long as both players adhere to a common run of the algorithm.

## 5 INFINITE VALUE ITERATION IN STOCHASTIC GAMES

So far we have presented an exact algorithm for computing Nash policy pairs in stochastic games with small state spaces, and a sparse sampling algorithm for computing approximately Nash policy pairs in large

---

**Algorithm InfiniteVI($T$):**
  **Initialization:** for all $s \in S$, $k \in 1, 2$:
    $Q_k[s, 0] \leftarrow M_k[s]$;
    $\pi_k(s) \leftarrow f_k(M_1[s], M_2[s])$;
  **Iteration** $t = 1, \ldots, T$: for all $s \in S$, $k \in \{1, 2\}$:
    For all pure strategies $i$ and $j$:
      $Q_k[s, t](i, j) \leftarrow M_k[s](i, j) + \gamma \sum_{s'} P(s'|s, i, j) v_f^k(Q_1[s', t-1], Q_2[s', t-1])$;
    $\pi_k(s) \leftarrow f_k(Q_1[s, t], Q_2[s, t])$;
  Return the policy pair $(\pi_1, \pi_2)$;

---

Figure 3: Algorithm **InfiniteVI** for computing a security pair in infinite-horizon discounted stochastic games. Here $f$ is an arbitrary security selection function, $f_k$ extracts the mixed strategy selected by $f$ for player $k$, and $v_f^k$ extracts the security level of the selected strategy to player $k$.

stochastic games. Both of these algorithms applied to the $T$-step average return setting. In this section, we examine the situation for the infinite horizon discounted return setting, and find some curious difference with the finite horizon case. In particular, we provide a small generalization of the classical result of Shapley on value iteration in the zero-sum case, but also provide a counterexample proving that the algorithm cannot (in a fairly strong sense) converge to Nash in the general-sum setting.

In Figure 3, we present a value iteration algorithm for the infinite-horizon discounted setting. This algorithm is quite similar to **FiniteVI**, except that the output policies $\pi_k(s)$ are now independent of time, and the discount factor is incorporated into the backup matrices computed at each step.

We now present a slight generalization of a classical result of Shapley (1953) on the convergence of infinite-horizon value iteration. Shapley proved the convergence in the case of zero-sum stochastic games, where the notions of Nash and security coincide. Here we give a more general analysis proving convergence to a security strategy in the general-sum setting.

**Theorem 5** *Let $G$ be any stochastic game, and let $f$ be any security selection function. Then as $T \to \infty$, the policy pair $\pi_k(s)$ output by **InfiniteVI**($T$) converges to a security pair for the infinite-horizon discounted stochastic game $G$ from any start state.*

A very natural question to ask about algorithm **InfiniteVI** is: if we allow the arbitrary security selection function $f$ to be replaced by an arbitrary *Nash* selection function, can we prove the generalization of Theorem 5 in which convergence to a security pair is replaced by convergence to a Nash pair? This would provide the infinite-horizon discounted analogue to Theorem 1 for the finite-horizon case.

In the full paper, we provide a counterexample proving that such a generalization is not possible. The counterexample is rather strong in several dimensions. First, it applies even to any *fixed* choice of Nash selection function. In other words, the difficulty does not lie in the generality of the Nash selection function, and particular choices or conditions on this function will not help. Second, the counterexample is such that there will be infinitely many time steps at which the policies currently computed by the algorithm are not even an approximate Nash pair. Thus, unlike the MDP setting, in stochastic games there is a significant difference in the convergence status of value iteration in the finite-horizon and infinite-horizon cases.